\documentclass[pra,twocolumn,superscriptaddress,showpacs,preprintnumbers,amsmath,amssymb]{revtex4-1}

\usepackage{amsmath}
\usepackage{graphicx}
\usepackage{dcolumn}
\usepackage{bm}
\usepackage{color}
\usepackage{enumerate}
\usepackage{mathrsfs}
\usepackage{url}
\usepackage{float}

\begin{document}
\bibliographystyle{apsrev4-1} 

\title{Efficient generation of many-body singlet states of spin-1 bosons in optical superlattices}

\author{Huanying Sun}
\affiliation{School of Physics and Technology, Wuhan University, Wuhan, Hubei 430072, China}

\author{Peng Xu}
\affiliation{School of Physics and Technology, Wuhan University, Wuhan, Hubei 430072, China}

\author{Han Pu}
\affiliation{Department of Physics and Astronomy, and Rice Center for Quantum Materials, Rice University, Houston, Texas 77251, USA}

\author{Wenxian Zhang}
\email[Corresponding email: ]{wxzhang@whu.edu.cn}
\affiliation{School of Physics and Technology, Wuhan University, Wuhan, Hubei 430072, China}

\date{\today}

\begin{abstract}
We propose an efficient stepwise adiabatic merging (SAM) method to generate many-body singlet states in antiferromagnetic spin-1 bosons in concatenated optical superlattices with isolated double-well arrays, by adiabatically ramping up the double-well bias. With an appropriate choice of bias sweeping rate and magnetic field, the SAM protocol predicts a fidelity as high as 90\% for a sixteen-body singlet state and even higher fidelities for smaller even-body singlet states. During their evolution, the spin-1 bosons exhibit interesting squeezing dynamics, manifested by an odd-even oscillation of the experimentally observable squeezing parameter. The generated many-body singlet states may find practical applications in precision measurement of magnetic field gradient and in quantum information processing.
\end{abstract}

\maketitle
\section{Introduction}
\label{sec:r1}
Many-body singlet state, theoretically predicted almost two decades ago, is the genuine quantum many-body ground state of an antiferromagnetic spin-1 Bose-Einstein condensate (BEC)~\cite{Ho1998,Law1998,Ohmi1998,Ueda2000,Ho2000}. It has attracted much attention due to its potential applications in improving gradient magnetometer~\cite{Hirano2013,Toth2013}, in realizing robust quantum state in decoherence-free subspace~\cite{Lidar1998,Stefanov2007,Stamper-Kurn2007,WangYao2010}, in understanding quantum magnetism in frustrated many-spin systems~\cite{PhysRevB.39.2344,Nature2010,Nature2011}, and in solving no-classical-solution problems in quantum information processing, such as $N$-strangers, secret sharing, and liar detection~\cite{PhysRevLett.89.100402}, etc. Nevertheless, due to the extremely strict requirement of an ultralow magnetic field in the order of microGauss, the many-body singlet state has yet been realized in experiments.

Much effort has been devoted experimentally to achieving an extremely weak magnetic field environment, which is a must to realize the quantum ground state of an antiferromagnetic spin-1 $^{23}$Na condensate. A remarkable advance has been made by Hirano's group, who suppressed their magnetic field within the range of 10 microGauss, using an active compensation technique in an expensive permalloy-metal-shielded room~\cite{Hirano2013}. However, even in such an ultralow magnetic field, it is not clear whether a robust antiferromagnetic ground state, in the form of the quantum many-body singlet state, can be reached in realistic experimental time scale.

Inspired by the merging of a few spin-1 bosons in a double well~\cite{PhysRevLett.101.090404,PhysRevLett.114.225302,PhysRevA.84.063636,PhysRevA.91.033608}, we propose in this work a stepwise adiabatic merging (SAM) protocol to generate the quantum many-body singlet state, by propagating adiabatically the antiferromagnetic spin-1 bosons in concatenated optical superlattices from an experimentally accessible initial state to the final (ground) quantum many-body singlet state. Briefly, we start with a Mott insulator state of spin-1 boson with single occupancy for each lattice site, with all the atoms optically pumped to the polar (i.e., $|F=1,m_F=0\rangle$) state, then slowly merge the nearest two lattice sites adiabatically along $x$ direction, and generate many two-body spin singlet states. Next, we merge again the adjacent two sites along $y$ and $z$ direction to obtain many eight-body singlet states. By further merging the next concatenation level optical lattice with longer wavelength, we obtain many sixty-four-body singlet states. For an $L$-level concatenation optical superlattices, the final singlet states are in principle $8^L$-body. The limitation on this SAM protocol is mainly the total evolution time, which is limited by the bosons' lifetime. Such a limitation can be alleviated by optimizing the protocol in an appropriate magnetic field. While in numerical simulations, we are also constrained by the computational power to a system size of 16 bosons. In this method, the final probability of the sixteen-body singlet state generated through the SAM protocol is above 90\% in our numerical simulations under current experimental conditions. The major advantage of our protocol is that it does not require an ultralow magnetic field, which is the limiting factor that prevents a direct realization of the many-body singlet state.
\par
The paper is organized as follows. In Sec.~\ref{sec:r2}, we describe the system of antiferromagnetic spin-1 atoms trapped in a double-well unit and the SAM protocol to generate many-body singlet states. In Sec.~\ref{sec:r3}, we discuss in detail three key ingredients of the SAM protocol, including the bias sweeping range, the applied magnetic field, and the evolution of the separated SAM steps. In Sec.~\ref{sec:r4}, we present a complete dynamical process to generate a sixteen-body singlet state with a fidelity as high as 90\%. The experimental observable, the generalized spin-squeezing parameter, is also discussed in this section. The conclusions are given in Sec.~\ref{sec:r6}. More details about the Hamiltonian, the numerical calculation, the oscillation of the fidelity, and the robustness of the SAM protocol are discussed in the Appendixes.

\section{Hamiltonian and SAM protocol}
\label{sec:r2}
We consider an ultracold dilute gas of bosonic atoms with hyperfine spin $F=1$ trapped in a concatenated optical superlattice with isolated double-wells in an external magnetic field along $z$ direction. Such a system is described  exactly by the standard Bose-Habburd model with spin degrees of freedom~\cite{PhysRevLett.81.3108,PhysRevA.68.063602,PhysRevA.84.063636}. Due to the conservation of the total particle number and total magnetic quantum number (setting as zero here), the linear Zeeman term does not affect the dynamics of the system, thus only the quadratic Zeeman effect is taken into account. The  Hamiltonian is
\begin{eqnarray}\label{eq:h1}
H &=& -J\sum_{\sigma=\pm 1,0} (\hat{L}_\sigma^\dagger\hat{R}_\sigma+\hat{R}_\sigma^\dagger\hat{L}_\sigma) +\frac{U_0}{2}\sum_{i=L,R}\hat{N}_i(\hat{N}_i-1) \nonumber \\
 &&+\frac{U_2}{2}\sum_{i=L,R}(\hat{\bf S}_i^2-2\hat{N}_i)+\varepsilon(\hat{N}_L-\hat{N}_R)+H_Z.
\end{eqnarray}
The first term describes the tunneling between wells in a double-well unit where $J=\int d^3{\bf{r}}\psi_L^*({\bf{r}})[-\hbar^2\nabla^2/(2M)+V({\bf{r}})]\psi_R({\bf{r}})$ depicts the tunneling amplitude with $\psi_{L(R)}(\bf{r})$ the wave function in left (right) well, $M$ the atom mass, and $V({\bf r})$ the effective potential for the double-well. The creation and annihilation operators $\hat{L}_\sigma^\dagger(\hat{R}_\sigma^\dagger)$ and $\hat{L}_\sigma(\hat{R}_\sigma)$ for the hyperfine spin state $\sigma\in\{-1,0,1\}$ in the left (right) well obey the canonical bosonic commutation relations. The intrawell density interaction is described by the repulsive $U_0>0$ term with $\hat{N}_L=\sum_\sigma \hat{L}_\sigma^\dagger\hat{L}_\sigma$ ($\hat{N}_R=\sum_\sigma \hat{R}_\sigma^\dagger\hat{R}_\sigma$) being the atom number operator in the left well (right well). The interaction strength is $U_{0,2}=c_{0,2}\int d^3{\bf{r}}|\psi_i({\bf{r}})|^4$, where $c_0=4\pi\hbar^2(a_0+2a_2)/(3M)$ and $c_2=4\pi\hbar^2(a_2-a_0)/(3M)$ with $a_{0,2}$ being respectively the $s$-wave scattering length of two colliding bosons with total angular momenta $0$ and 2~\cite{Ho1998,Ohmi1998,Law1998}. The intrawell antiferromagnetic spin exchange interaction is described by $U_2>0$ term, where $\hat{\bf S}_L=\sum_{\sigma'\sigma}{L}_\sigma^\dagger{\bf F}_{\sigma\sigma'}{L}_{\sigma'}$ ($\hat{\bf S}_R=\sum_{\sigma'\sigma}{R}_\sigma^\dagger {\bf F}_{\sigma\sigma'}{R}_{\sigma'}$) is the total spin in the left (right) well with ${\bf F}_{\sigma\sigma'}$ being the standard spin-1 matrices. The term with $\varepsilon$ is the bias between the left and right well. The quadratic Zeeman energy, $H_Z=q\sum_{i=L,R}(\hat N_{i,+1}+\hat N_{i,-1})$, is either for a magnetic field, $q=q_0B^2$ with $q_0 = 277$ Hz/G$^2$ for $^{23}$Na atoms~\cite{Nature.396}, or a negative quadratic Zeeman shift generated by a microwave driving field~\cite{PhysRevA.73.041602,PhysRevLett.107.195306,PhysRevA.79.043631,PhysRevA.89.023608,SM}.

\begin{figure}
\centering
\includegraphics[width=3.25in]{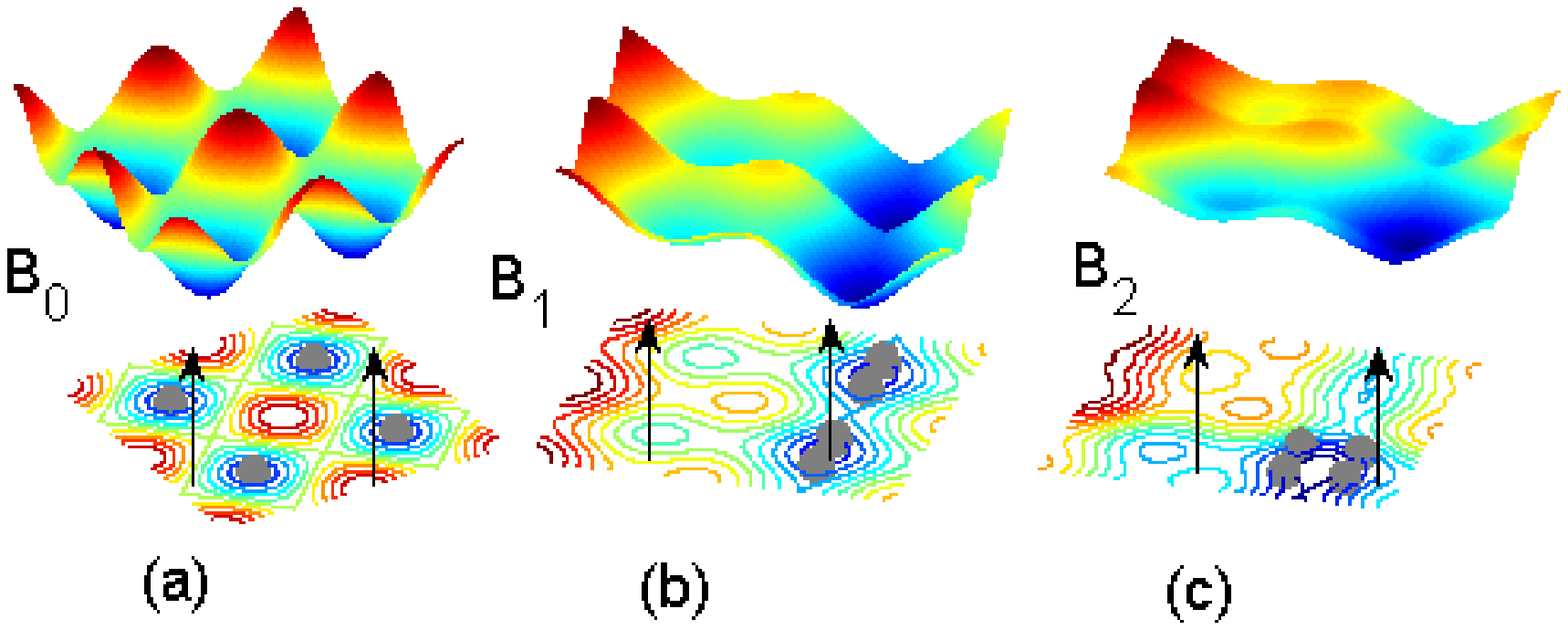}
\includegraphics[width=3in]{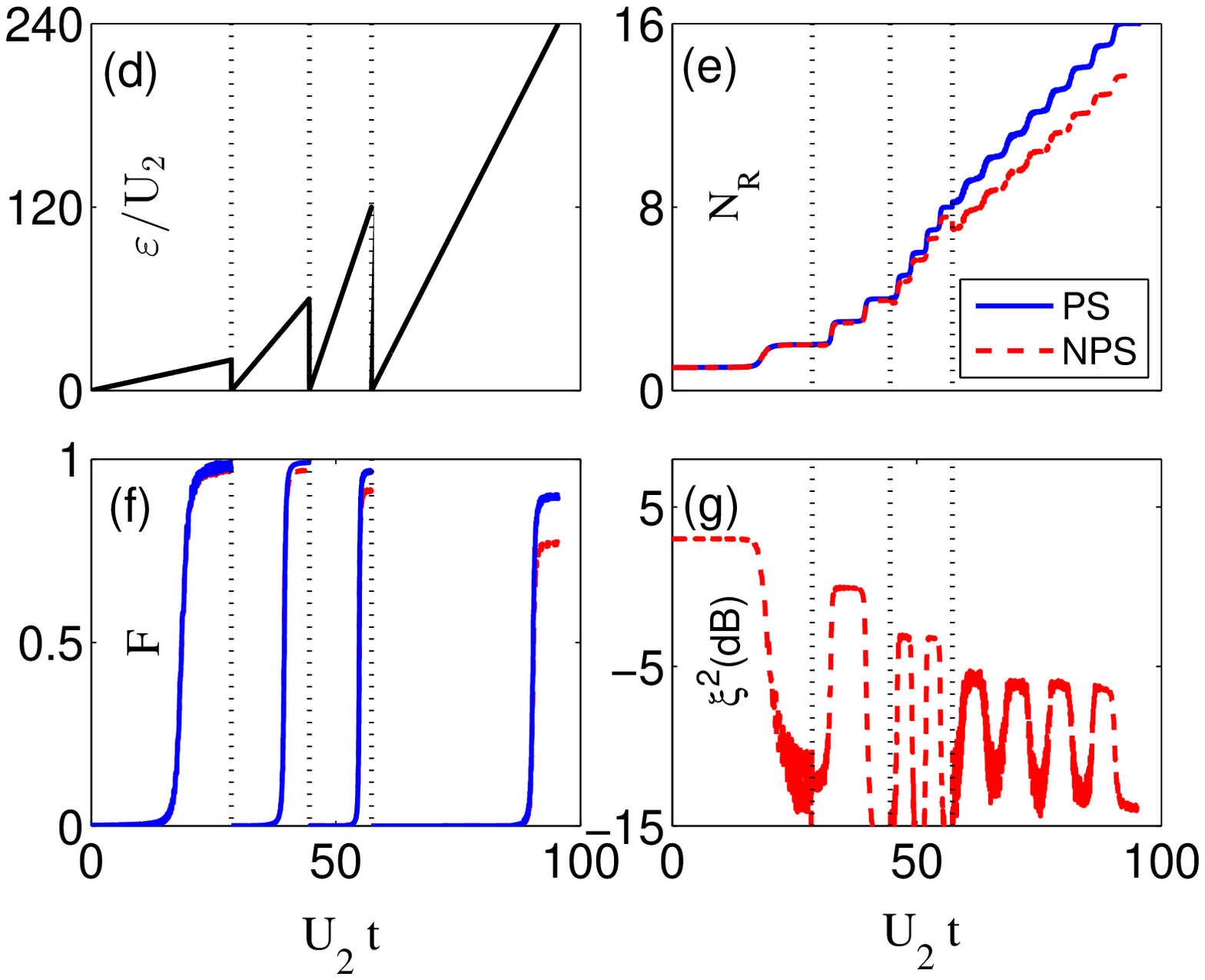}
\caption{\label{fig:1} (Color online.) Schematic of generation of many-body singlet states in an antiferromagnetic spin-1 BEC. (a)---(c) Illustrations of the four-body singlet state generation. (a) Atoms are placed in an optical lattice with single occupation in a strong magnetic field $B_0$. The spin state of the atoms is a polar state, $|F=1, m_F=0\rangle$. (b) The two-body singlet state is produced by adiabatically ramping up the left wells in a magnetic field $B_1$. (c) The four-body singlet state is generated by adiabatically ramping up the back wells. Higher-many-body singlet states are generated similarly by employing concatenated optical superlattice. (d) The bias and (e) lower-well occupation numbers dependence on time for the generation of the sixteen-body singlet state. (f) Fidelities  and (g) the generalized spin-squeezing parameter of the many-body singlet state during the evolution. The blue solid and red dashed lines, respectively, denote the results with and without postselection. Vertical dotted lines in (d)---(g) denote the connecting point of the adjacent SAM steps.}
\end{figure}

The SAM protocol is a successive process, as shown in Fig.~\ref{fig:1}(a-c). The system starts with a singly occupied polar state for antiferromagnetically interacting spin-1 bosons. By adiabatically ramping up the bias within each isolated double-well unit, say, along the $x$ direction (from Fig.~\ref{fig:1}(a) to (b)), two atoms are merged together in the lower well and form a two-body singlet state. Clearly, the only dynamical parameter is the bias in this merging process. Next, repeating the step along the $y$ direction (from Fig.~\ref{fig:1}(b) to (c)), four atoms are merged in a lower well and a four-body singlet state is generated. Continuing this adiabatic merging again along the $z$ direction, we obtain eight-body singlet states. Obviously, in order to generate larger than eight-body singlet states, a concatenated double-well optical superlattice with multiple light wave lengths is required by the SAM protocol.

\section{Three key ingredients}
\label{sec:r3}
\begin{figure}
\centering
\includegraphics[width=3.25in]{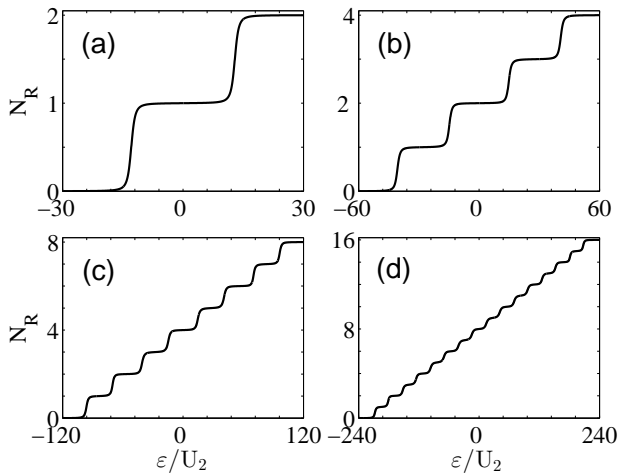}
\caption{\label{fig:2} Dependence of the lower-well occupation number on the bias for the generation of (a) two-body, (b) four-body, (c) eight-body, and (d) sixteen-body singlet states, according to the SAM protocol. Parameters are the same as in Fig.~\ref{fig:1} except that here $q=0$. In the SAM protocol, $\varepsilon \geq 0$ is required.}
\end{figure}

Before we present the complete evolution of the spin-1 bosons under the SAM protocol, let us discuss the three key ingredients separately.  Firstly, we need to determine the sweeping range of the double-well bias for each SAM step. Such a range is determined by exploring the dependence of $N_R$, the occupation number of the lower well which we refer as ``right" well hereafter, on the double-well bias $\varepsilon$ for the ground state of the Hamiltonian Eq.~(\ref{eq:h1}). The results are shown in Fig.~\ref{fig:2} for the merging of $2 \sim 16$ atoms. It can be seen that $N_R$ increases in steps of one as the bias increases. Such a single atom tunneling is due to the strong intrawell repulsion, $U_0\gg J$. This kind of behavior has been investigate in theory~\cite{PhysRevA.78.031601,PhysRevA.78.023606,PhysRevA.83.023608} and confirmed in experiment~\cite{ PhysRevLett.101.090404}. The atom number $N_R$ eventually reaches its maximum at a large bias $\varepsilon_f$, thus the range is $[0,\varepsilon_f]$. The merge of larger number of atoms requires a larger bias range, roughly in a linearly increasing form. This in fact manifests the linear relation between the chemical potential and the number of atoms in a double well, $\varepsilon_f = dE_N/dN \approx U_0N/2 $, where the energy is $E_N \approx (U_0/2)N(N-1) - E_{b}$ with $E_b \approx (U_0/4)N(N-1)$ being the energy of the system in a balanced double well $\varepsilon=0$.

\begin{figure}
\centering
\includegraphics[width=3.25in]{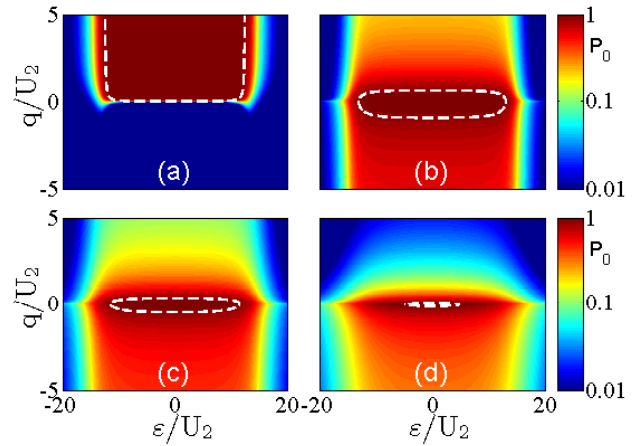}
\caption{\label{fig:3} (Color online.) (a) Polar product state probability, (b) two-body singlet product state probability, (c) four-body singlet product state probability, and (d) eight-body singlet product state probability in a double well. The white dashed lines mark the probability of 90\%. Parameters are the same as in Fig.~\ref{fig:1}. During the generation of a many-body singlet state, we choose a magnetic field and an initial bias within the region enclosed by the white dashed lines ($P_0>90\%$).}
\end{figure}

Secondly, we determine the range of the magnetic field, in which the ground state is close to the expected initial state of each SAM step. According to the protocol, the ideal initial state $|\psi_E \rangle$ is the product state of the left and right well with exactly the same quantum state, e.g., a polar product state for total 2 atoms, a $N/2$-body singlet product state for $N$ atoms, and so on. To evaluate the efficiency of preparing the initial state, we define a probability $P_0(\varepsilon,q) = |\langle \psi_G|\psi_E\rangle|^2$ where $|\psi_G\rangle$ is the ground state of the system at a finite bias $\varepsilon$ and quadratic Zeeman energy $q$ ($U_{0,2}$ and $J$ are given).

We plot the probability $P_0$ in Fig.~\ref{fig:3}. For $N=2$ atoms, the probability $P_0$ increases as $q$ increases in low bias $|\varepsilon|$ region. Thus, the expected initial state $\psi_E$ can be prepared in a large magnetic field at zero (low) bias. For $N>2$ and even number of atoms, the probability $P_0$ is high around the central region, i.e., low bias ($\varepsilon$) and low quadratic Zeeman energy ($|q|$) region. In order to generate a final many-body singlet state with a probability higher than $90\%$ with the SAM protocol, we are limited to choose a set of $\varepsilon$ and $q$ within the enclosed region marked by the white dashed line (where $P_0>90\%$). As shown in the figure, the largest quadratic Zeeman energy on the white dashed line is $q/U_2 = 0.64$, $0.34$, and $0.054$ in the panels 3(b), 3(c), and 3(d), respectively. The enclosed region becomes smaller and smaller as $N$ increases, indicating that the generation of larger many-body singlet states becomes more and more challenging~\cite{egnote}. This is the main reason why the many-body singlet state has not been observed experimentally though it has been predicted theoretically for almost two decades~\cite{Law1998,Ho2000,Ueda2000}.

Thirdly, we investigate the performance of the separated SAM steps for a fixed even number of atoms in a double well unit, by assuming an ideal initial state $|\psi_E \rangle$~\cite{initnote}. During the evolution, we adiabatically ramp up the bias at a constant rate from zero to a final value $\varepsilon_f$. The evolution of the system is monitored by two observables, the number of atoms in the right well $N_R$ and the fidelity $F = |\langle\Psi(t)|\Psi_S\rangle|^2$ with $|\Psi(t) \rangle$ being the state vector at time $t$ and $|\Psi_S \rangle =|N,S=0,M_S=0\rangle  $ the targeted $N$-body singlet state.

\begin{figure}
\centering
\includegraphics[width=3.25in]{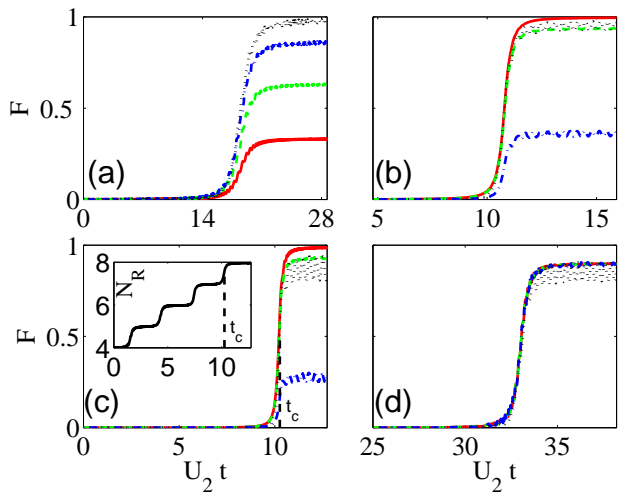}
\caption{\label{fig:4} (Color online.) Time dependence of the fidelity of (a) two-body, (b) four-body, (c) eight-body, and (d) sixteen-body singlet state. The quadratic Zeeman energies during the evolution are $q/U_2 = 0$ (red solid lines), $0.0069$ (green dashed lines), $0.0277$ (blue dash-dotted lines), and $0.1108$ (black dotted lines). The inset in (c) shows typical time dependence of the lower-well occupation number $N_R$. Other parameters are the same as in Fig.~\ref{fig:1}. The vertical dashed lines denote the time $t_c$ where the $N_R$ approaches to the total number of atoms $N$.}
\end{figure}

We illustrate the evolution of the generation of the many-body singlet states according to the SAM protocol in Fig.~\ref{fig:4} for atom number $N=2, 4, 8$, and $16$ at four chosen quadratic Zeeman energies. The fidelity increases sharply from zero to an almost constant at a time $t_c$, which coincides with the time when the right well atoms $N_R$ approaches to the total number of atoms $N$ [see Fig.~\ref{fig:4}(c)]. The highest fidelity for each $N$ is above 90\% among the four selected quadratic Zeeman energy within an appropriate evolution time. From Fig.~\ref{fig:4}, we also observe that the fidelity at the end of evolution exhibits oscillations with the increasing of the quadratic Zeeman energy, indicating the existence of an optimal magnetic field in a realistic experimental situation \cite{SM}.

\section{Efficient generation of singlet states}
\label{sec:r4}

Finally, to present a complete view, we carry out a continuous evolution process to efficiently generate a sixteen-body singlet state from a singly occupied polar state in a concatenated optical superlattice, according to the SAM protocol shown in Fig.~\ref{fig:1}. The sweeping rate of the bias is a constant during each step. However, the sweeping rate is adjusted for different steps (see the lines in Fig.~\ref{fig:1}(d)), in order to limit the total evolution time in an experimentally accessible regime. The total number of atoms of the generated singlet state is in principle doubled after the merging of each step of the SAM protocol. In our simulation, the initial state of the next step is manually set as the product of the final state in the lower well and its copy. Such an operation is nonunitary so that the total number of atoms $N$ as well as the atoms in the lower well $N_R$ in fact decrease at this connecting point, as shown in Fig.~\ref{fig:1}(e). However, the product state at the beginning of each SAM protocol is important to reduce the computational basis from an exponential increasing $3^N$ to a much slower way, see Table~\ref{table:1} in Appendix~\ref{app:c}.

The fidelities of the many-body singlet state are presented in Fig.~\ref{fig:1}(f). Clearly, the SAM protocol is efficient to produce the many-body singlet states with high fidelity. As a trend, the larger the size of the many-body singlet state, the lower the final fidelity. Such a decline of the fidelity is caused by two sources, the nonadiabaticity during the evolution and the atom loss between two adjacent SAM steps. The nonadiabaticity caused fidelity declination may be prevented by sweeping the bias with a slower rate or by employing nonlinear sweeping function, such as the  shortcut to adiabatic passage ~\cite{XiChen2010Shortcut}. The atom loss caused fidelity dropping can be improved by utilizing the experimental technique of postselection, i.e., we only take into account of the results having the number of atoms in the lower well $N_R$ exactly equal to $2^n$ ($N_R = 16$ here). As shown in our calculation (Fig.~\ref{fig:1}(e, f)), the final sixteen-body singlet fidelity jumps from 77.5\% to a value above 90\% by employing the postselection method. Here, the parameters are $U_0/U_2=27.8$ and $J/U_2=0.694$. The quadratic Zeeman energies are $q/U_2$ = 0.1108 ($U_2t < 28.6$) and 0.0069 ($U_2t > 28.6$) for the generation of two-body singlet state and the generation of other higher many-body singlet states, respectively. To simulate the magnetic field fluctuation which is inevitable in practice, we included a white noise with amplitude 1 mG and found that it has very small effect on the performance of our protocol \cite{SM}. To control the fluctuation of the magnetic field strength to within 1 mG is very much feasible in most labs. This clearly demonstrates that our protocol does not require an ultralow magnetic field strength on the order of microGauss.

It is a big challenge to directly detect the many-body singlet state fidelity in an antiferromagnetic spin-1 BEC experiment. To circumvent this obstacle, we propose to monitor the generation of the many-body singlet state with the generalized spin-squeezing parameter, which is used to estimate the entanglement level of a quantum state~\cite{PhysRevLett.99.250405,1367-2630-12-5-053007,PhysRevLett.107.240502,PhysRevLett.113.093601},
$$
\xi^2=\frac 1 {FN} {\sum_{\alpha=x,y,z} (\Delta \hat{S}_\alpha)^2}
$$
where ${\hat {\bf S}}$ is the total spin angular moment, $F=1$ and $N$ the total number of atoms. A spin state is squeezed if $\xi^2<1$, compared to a coherent spin state with $\xi^2 =1$. For a perfect many-body singlet state, obviously $\xi^2 = 0$ since $\langle {\bf S}\rangle=0$ and $\langle {\bf S}^2 \rangle=0$~\cite{PhysRevLett.113.093601}.

As shown in Fig.~\ref{fig:1}(g), the squeezing parameter decreases as time evolves and suddenly drops to a value below -10 dB during the first SAM step, manifesting the generation of the two-body singlet state. At the beginning of the second SAM step, the generalized squeezing parameter increases since the addition of the third atom breaks the many-body singlet state, due to the fact that a many-body singlet state requires an even $N$ in identical spin-1 bosons~\cite{Law1998}. Similar to the first SAM step, the second sudden dropping of the squeezing parameter indicates the production of the four-body singlet state. This odd-even oscillation of the squeezing parameter continues in the later SAM steps and offers an excellent experimental witness of the even-body spin singlets. This is in contrast to the usual detection of the number fluctuation of each component, which is large but changes little during the evolution~\cite{Law1998, Ho2000}. At the end of the fourth SAM step, the squeezing parameter is still below -10 dB, indicating a high efficiency of the generation of the sixteen-body singlet state.

The SAM protocol is practical to implement under current experimental conditions. In $^{23}$Na spin-1 boson experiments, a typical value of $U_2$ is estimated as 50 Hz. The total evolution time is thus about $1.9$ s in Fig.~\ref{fig:1} (we set $\hbar=1$ in our calculations). The qudratic Zeeman energies are $5.5$ Hz for $N=2$ and $0.35$ Hz for later steps, which correspond to magnetic fields of $141.4$ mG and $35.4$ mG, respectively, which are easily accessible in current ultracold atomic gases experiments. Obviously, due to the conservation of the magnetization which cancels the linear Zeeman effect, the magnetic fields in the SAM protocols are much larger than the previous estimations of 10$^{-7}$ G where the global ground state is considered~\cite{Ueda2000}.

\section{Conclusion}
\label{sec:r6}
In conclusion, we have proposed a stepwise adiabatic merging protocol to generate efficiently the long-sought many-body spin singlet state in antiferromagnetic spin-1 bosons in a concatenated optical superlattice. Our numerical simulations show that the generation efficiency of a sixteen-body singlet state is as high as 90\% under the current experimental conditions. The evolution of the SAM protocol can be witnessed conveniently by the generalized spin squeezing parameter, which exhibits large amplitude odd-even oscillations. The generated many-body spin singlet states provide a stepstone to reach the quantum limit gradient magnetometer with spin-1 bosons and to solve the famous problems in quantum information science~\cite{Toth2013,PhysRevLett.89.100402}.

\begin{acknowledgements}
We thank Y. Liu for helpful discussions and providing experimental parameters. HS, PX, and WZ are grateful to Beijing CSRC for hospitality in the early stage of this project and are supported by the National Natural Science Foundation of China under Grant Nos. 11574239, 11275139, and 11647312, and the National Basic Research Program of China (Grant No. 2013CB922003). HP is supported by the US NSF and the Welch Foundation (Grant No. C-1669).
\end{acknowledgements}

\appendix

\section{Hamiltonian of  the system}
\label{app:H}
We consider a dilute gas of bosonic atoms with hyperfine spin $F = 1$ trapped in a concatenated optical superlattice with isolated double-wells in an external magnetic field along z direction. This system conserves the total particle number ($N$) and total magnetic quantum number
(setting as $M=0$ here), thus only the quadratic Zeeman effect is taken into account. The Hamiltonian of the system, Eq. (1) in main text, is expanded as~\cite{PhysRevLett.81.3108,PhysRevA.68.063602,PhysRevA.84.063636}
\begin{widetext}
\begin{eqnarray}\label{eq:h1}
H&=&\frac{U_0}{2}[\hat{N}_L(\hat{N}_L-1)+\hat{N}_R(\hat{N}_R-1)]-J(\hat{a}_{L-1}^\dagger\hat{a}_{R-1}
+\hat{a}_{R-1}^\dagger\hat{a}_{L-1}+\hat{a}_{L0}^\dagger\hat{a}_{R0}+\hat{a}_{R0}^\dagger\hat{a}_{L0}
+\hat{a}_{L1}^\dagger\hat{a}_{R1}+\hat{a}_{R1}^\dagger\hat{a}_{L1}) \nonumber \\
&+&\varepsilon(\hat{N}_L-\hat{N}_R)+\frac{U_2}{2}(\hat{{\bf S}}_L^2-2\hat{N}_L)+
\frac{U_2}{2}(\hat{{\bf S}}_R^2-2\hat{N}_R)+q (\hat{N}_{L1}+\hat{N}_{L-1}+\hat{N}_{R1}+\hat{N}_{R-1}).
\end{eqnarray}
The coefficients, $U_0, J, U_2, q$ have been described in the main text. $\hat{N}_i=\hat{a}_{i1}^\dagger\hat{a}_{i1}+\hat{a}_{i0}^\dagger\hat{a}_{i0}+\hat{a}_{i-1}^\dagger\hat{a}_{i-1}$ is the atom number operator in the $i$th ($i=L, R$) well. The components of the spin-1 vector $\hat{\bf{S}}$ are written as creation and annihilation operators,
\begin{eqnarray}
\hat{S}_x &=& \frac{1}{\sqrt{2}}(\hat{a}_1^\dagger\hat{a}_0+\hat{a}_0^\dagger\hat{a}_{-1}+ \hat{a}_0^\dagger\hat{a}_1+\hat{a}_{-1}^\dagger\hat{a}_0), \nonumber \\
\hat{S}_y &=& \frac{i}{\sqrt{2}}(-\hat{a}_1^\dagger\hat{a}_0-\hat{a}_0^\dagger\hat{a}_{-1}+ \hat{a}_0^\dagger\hat{a}_1+\hat{a}_{-1}^\dagger\hat{a}_0), \nonumber \\
\hat{S}_z &=& (\hat{a}_1^\dagger\hat{a}_1-\hat{a}_{-1}^\dagger\hat{a}_{-1}).
\end{eqnarray}
Based on these equations, the $U_2$ terms in Eq.~(\ref{eq:h1}) becomes
\begin{eqnarray}\label{eq:h2}
\frac{U_2}{2}(\hat{\bf S}_i^2-2\hat{N}_i) &=& \frac{U_2}{2}[\hat{N}_{i1}(\hat{N}_{i1}-1) +\hat{N}_{i-1}(\hat{N}_{i-1}-1)+2\hat{N}_{i1}\hat{N}_{i0}
+2\hat{N}_{i0}\hat{N}_{i-1}-2\hat{N}_{i1}\hat{N}_{i-1} \nonumber \\
&& +2(\hat{a}_{i0}^\dagger)^2\hat{a}_{i1}\hat{a}_{i-1} +2\hat{a}_{i1}^\dagger\hat{a}_{i-1}^\dagger(\hat{a}_{i0})^2],
\end{eqnarray}
\end{widetext}
where $\hat{a}_{i\sigma}^\dagger$$(\hat{a}_{i\sigma})$ is the creation (annihilation) operator of the hyperfine state with $\sigma\in\{-1,0,1\}$.

\section{Construction of many-body singlet states}

The many-body spin singlet state, $|N,S=0,M_S=0\rangle$, is consisted of $N$ particles with total angular momentum quantum number $S=0$. For a two-body singlet state, the $|2,S=0,M_S=0\rangle$ can be theoretically produced by acting the singlet pair operator $\hat{A}^\dagger=[(\hat{a}_0^\dagger)^2-2\hat{a}_1^\dagger\hat{a}_{-1}^\dagger]/\sqrt{3}$ on the vacuum state $|{\rm vac}\rangle$ with a following normalization
\begin{eqnarray*}
|2,S=0,M_S=0\rangle &=& \sqrt{\frac{1}{3}}\;|0,2,0\rangle - \sqrt{\frac{2}{3}}\;|1,0,1\rangle
\end{eqnarray*}
where the state $|k,N-2k,k\rangle$ for $k=0,1$ denotes the basis of Fock states. For even $N$ atoms, the singlet state $|N,S=0,M_S=0\rangle$ is constructed by acting the singlet pair operator consequently~\cite{Law1998,Ho2000,Ueda2000},
$|N,S=0,M_S=0\rangle=(\hat{A}^\dagger)^{\frac{N}{2}}|{\rm vac}\rangle$. The state after the normalization becomes
\begin{eqnarray}\label{eq:h3}
|N,S=0,M_S=0\rangle=\sum_{k=0}^{N/2}A_k|k,N-2k,k\rangle
\end{eqnarray}
with the amplitudes $A_k$ obeying the following recursion relation
\begin{eqnarray}\label{eq:h4}
A_k=-\sqrt{\frac{N-2k+2}{N-2k+1}}A_{k-1}.
\end{eqnarray}
As an example, the four-body singlet state is
\begin{eqnarray*}
|4,S=0,M_S=0\rangle&=&\sqrt{\frac{1}{5}}|0,4,0\rangle-\sqrt{\frac{4}{15}}|1,2,1\rangle \\ &+&\sqrt{\frac{8}{15}}|2,0,2\rangle.
\end{eqnarray*}

\section{Diagonalization of the Hamiltonian}
\label{app:c}

We work in the computational basis of the Fock space in an isolated double well unit, $|N_{L1}, N_{L0}, N_{L-1}; N_{R1}, N_{R0}, N_{R-1}\rangle$. We set the parameters $U_2=1$, and $U_0/U_2=c_0/c_2\approx 27.78$ for $^{23}$Na spin-1 Bose-Einstein condensates and $J/U_2=U_0/40 \approx 0.694$ throughout the Letter \cite{PhysRevA.63.012710,PhysRevLett.114.225302,PhysRevA.83.042704}. The number of basis states $K$ of $N$ atoms  in a double well increase roughly in an exponential form, as shown in TABLE.\ref{table:1}.
\begin{table}
\caption{\label{table:1} Number of basis state $K$ increases roughly exponentially with the number of atoms $N$ in a double well unit system.}
\begin{tabular}
{m{1cm}<{\centering}|m{1cm}<{\centering}m{1cm}<{\centering}m{1cm}<{\centering}p{1cm} <{\centering}p{1cm}<{\centering}}
\hline \hline
$N$ &2 & 4 & 8 & 16 & 32 \rule{0pt}{0.45cm}\\
\hline
$K$ &7 & 26 &155& 1365& 15657 \rule{0pt}{0.45cm}\\
\hline \hline
\end{tabular}
\end{table}

In the main text, Fig.~2 is obtained by searching for the ground state via diagonalizing the Hamiltonian with zero quadratic Zeeman energy for different potential bias $\varepsilon$. Then we calculate the expectation value of the atom number in the right well $N_R = \langle {\hat N}_R\rangle$. Similarly, the probability $P_0$ in Fig. 3 in the main text is obtained by calculating the overlap of the ideal many-body singlet product state (the polar product state for $N=2$) with the found ground state at different quadratic Zeeman energy $q$ and different potential bias $\varepsilon$.

For the SAM protocol shown in Fig.~\ref{fig:1}, the minimum gap $\Delta$ between the instantaneous first excited state and the ground state are $0.07U_2, 0.05U_2, 0.13U_2$, and $0.27U_2$ for the SAM steps during the generation of the 2-body, 4-body, 8-body, and 16-body singlet states, respectively. Multiplied by the  evolution time $T$ of each SAM step, we find $\Delta\; T \gtrsim 1$. This relation roughly satisfies the adiabatic condition.

\section{Oscillations of fidelity}

\begin{figure}
\centering
\includegraphics[width=3.25in]{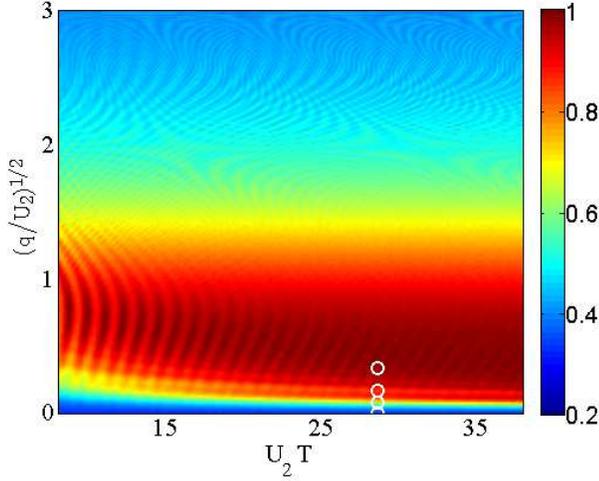}
\caption{\label{fig:5} (Color online.) Final fidelity of the two-body singlet state versus the total evolution time and the quadratic Zeeman energy. The four white circles correspond to the four simulations in Fig. 4(a) in the main text. Clearly, there is a wide high-fidelity region.}
\end{figure}

\begin{figure}
\centering
\includegraphics[width=3.25in]{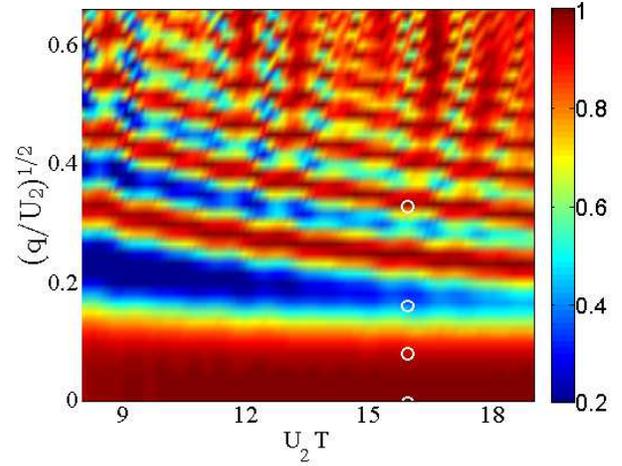}
\caption{\label{fig:6} (Color online.) Same as in Fig.~\ref{fig:5} except for four-body singlet state. The four white circles correspond to the four simulations in Fig. 4(b) in the main text. Different from the two-body singlet state case, there are many high-fidelity regions for the four-body case, with many oscillations in the magnetic field region we consider.}
\end{figure}

We observe in Fig.~4 some signatures of oscillatory behavior of the fidelity for the four chosen values of the quadratic Zeeman energy. A more systematic investigation of the fidelity dependence on the total evolution time and the quadratic Zeeman energy are presented in Fig.~\ref{fig:5} and Fig.~\ref{fig:6} for $N=2$ and $N=4$ atoms, respectively. The fidelities are the final value at the end of the evolution with the given quadratic Zeeman energy. The bias sweeps linearly from zero to $\varepsilon_f = 20 U_2$ for $N=2$ atoms and to $60 U_2$ for $N=4$ atoms.

From Fig.~\ref{fig:5} we observe an oscillation with the quadratic Zeeman energy for a fixed evolution time. As shown in the figure, the fidelity is not very sensitive to the change of the total evolution time. While for the quadratic Zeeman energy, there exists a pretty large optimal region around 0.25 where the fidelity is high.

In Fig.~\ref{fig:6} there is a high-fidelity region with weak dependence on the total evolution time near zero quadratic Zeeman energy. As the quadratic Zeeman energy increases, the fidelity shows many oscillations and there are several high-fidelity bands. These results indicate that one needs to set wisely in an experiment the quadratic Zeeman energy and the evolution time, in order to reach a high fidelity of many-body singlet state.


\section{Robustness of the SAM protocol}

\begin{figure}
\centering
\includegraphics[width=3.25in]{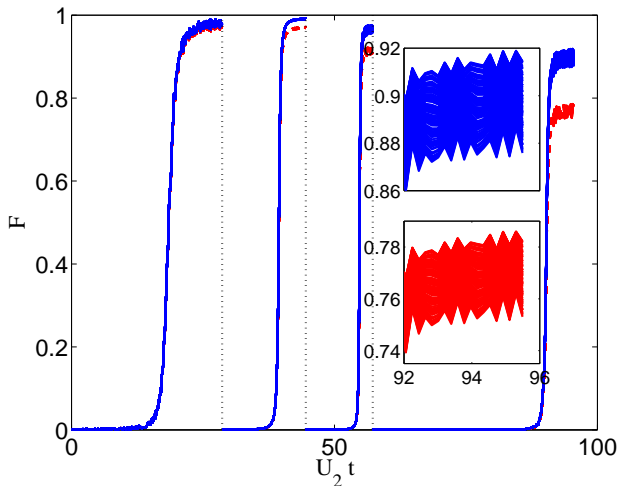}
\caption{\label{fig:7} (Color online.) Time dependence of the fidelities under the SAM protocol for 100 realizations of random magnetic field within [140.4,142.4] mG for the two-body singlet state and [34.4,36.4] mG for higher-body singlet states. Red dashed lines represent the results without post-selection and blue solid lines with post-selection. The insets illustrate the results near the end of evolution.}
\end{figure}
\begin{figure}[h]
\centering
\includegraphics[width=3.25in]{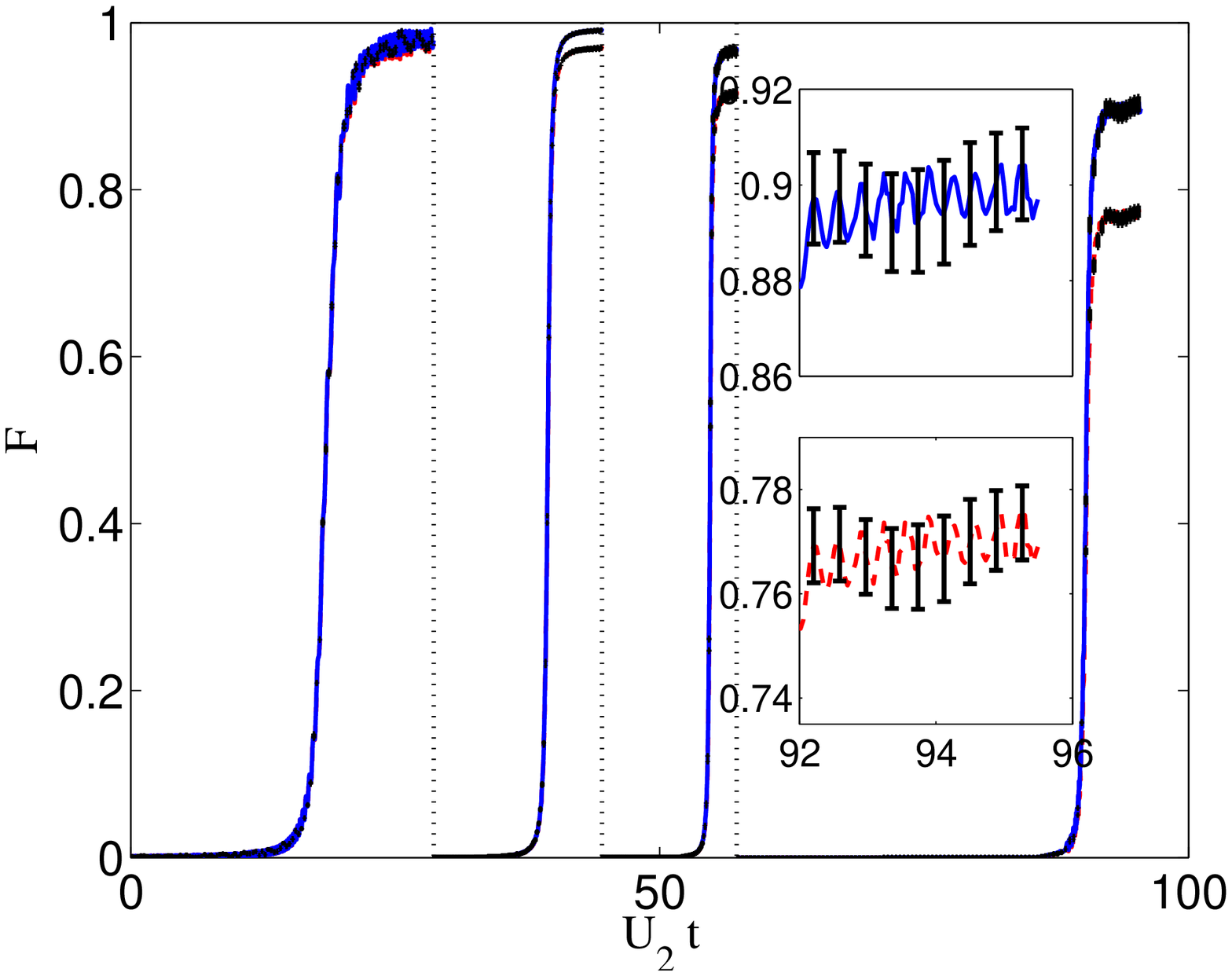}
\caption{\label{fig:8} (Color online.) Same as Fig.~\ref{fig:7} except that the average and the error bars of the 100 realizations. The results show that the SAM protocol is quite robust against the magnetic field fluctuation in experiments.}
\end{figure}

In experiments, the magnetic field may fluctuate from shot to shot. To test the robustness of the proposed SAM protocol, we assume uniformly distributed random magnetic fields around its averages, 141.4 mG ($q/U_2\approx 0.1108$ for $^{23}$Na atoms with $U_2 = 50$ Hz) for the two-body singlet state and 35.4 mG ($q/U_2\approx 0.0069$) for the higher-body singlet state. The magnetic field fluctuation is within 1 mG, which is easily realized with current experimental techniques. The results for 100 realizations are shown in Fig.~\ref{fig:7} and the averages and typical error bars are shown in Fig.~\ref{fig:8}. Clearly, the final fidelities of the sixteen-body singlet state only fluctuate in a small range, indicating the robustness of the SAM protocol.


\begin{thebibliography}{39}
\bibitem [{egn()}]{egnote}%
  \BibitemOpen
  \href@noop {} {}\bibinfo {note} {For example, the quadratic Zeeman energy
  $q/U_2<0.36$ is required to satisfy $P_0>90$\% at $\varepsilon=0$ in
  Fig.~\ref{fig:3}(c), but it drops to $q/U_2<0.054$ in
  Fig.~\ref{fig:3}(d).}\BibitemShut {Stop}%
\bibitem [{ini()}]{initnote}%
  \BibitemOpen
  \href@noop {} {}\bibinfo {note} {For an even $N$ atoms system, the initial
  state is set as a product state of two singlet states with $N/2$ atoms each.
  While for $N=2$ atoms, the initial state is a product state of two polar
  states with a single atom in each well.}\BibitemShut {Stop}%
\end{thebibliography}
%

\end{document}